\documentclass[a4paper]{jpconf}
\usepackage{graphicx}
\begin{document}
\title{{\underline {\bf MIMAC}}: A micro-tpc matrix for  dark matter directional detection}

\author{}

\author{D.~Santos, J.~Billard$^1$,  G.~Bosson, J.L. ~Bouly, O.~Bourrion, \\C. ~Fourel, O.~Guillaudin, J.~Lamblin, J.F.~Muraz, F.~Mayet, J.P.~Richer, Q.~Riffard}
\address{  LPSC, Universite Joseph Fourier Grenoble 1, CNRS/IN2P3, Institut Polytechnique de Grenoble}
\address{$^1$ Present address:  Department of Physics and Kavli Institute for Astrophysics and Space Research, Massachusetts Institute of Technology, Cambridge, MA 02139, USA, }
%\address[MIT]{Department of Physics, Massachusetts Institute of Technology, Cambridge, MA 02139, USA}
%\address[MITK]{MIT Kavli Institute for Astrophysics and Space Research, Massachusetts Institute of Technology, Cambridge, MA 02139, USA}

\author{E.~Ferrer, I.~Giomataris, F.J.~Iguaz$^2$}
\address{	IRFU,CEA Saclay, 91191 Gif-sur-Yvette cedex}
\address{$^2$ Present address: Laboratorio de F\' {i}sica Nuclear y Astropart\' {i}culas, Universidad de Zaragoza, Spain}

%\author{J. ~Busto, J.~Brunner, D.~Fouchez}
%\address{	CPPM, Marseille}

\author{L.~Lebreton, D.~Maire}
\address{	LMDN, IRSN Cadarache, 13115 Saint-Paul-Lez-Durance}

\ead{Daniel.Santos@lpsc.in2p3.fr}

\begin{abstract}

 The dark matter directional detection opens a new field in cosmology bringing the possibility to build a map of nuclear recoils that would be able to explore the galactic dark matter halo giving access to a particle characterization of such matter and the shape of the halo.
The MIMAC (MIcro-tpc MAtrix of Chambers) collaboration has developed in the last years an original prototype detector based on the direct coupling of large pixelized micromegas with a  devoted fast self-triggered electronics showing the feasibility of a new generation of directional detectors. The discovery potential of this search strategy is discussed and illustrated. In June 2012, the first bi-chamber prototype has been installed at Modane Underground Laboratory (LSM)  and the first underground background events,  the gain stability and calibration are shown.
\end{abstract}

\section{Introduction}

Cosmological and astrophysical observations converge to a standard cosmological model requiring a new kind of particle. The most favourable candidate is a stable weakly interacting massive particle (WIMP). In the context of direct detection of these new particles an alternative, complementary and in any case needed strategy is the development of detectors providing an unambiguous positive WIMP signal. Indeed, directional detection gives a new degree of freedom to discriminate statistically the WIMP interactions from  neutrons, the particles that produce the same expected signal, giving in addition a clear signature for events produced by particles in our galactic halo. This can be achieved by searching for a correlation of the WIMP signal with the solar motion around the galactic centre, observed as a direction dependence of the WIMP stream \cite{spergel}, coming from (l = 90$^\circ$, b = 0$^\circ$) in galactic coordinates, which happens to be roughly in the direction of the constellation Cygnus. The background events, coming from gamma rays and neutrons produced in the atmosphere or in the rock should follow the rotation of the Earth,  isotropic in galactic coordinates and easy to discriminate from those coming from the Cygnus constellation direction.
A dedicated statistical study with simulated data analysis has shown that even a low-exposure, directional detector could allow a high significance discovery of galactic Dark Matter even with a  background contamination or to a robust and competitive exclusion curve \cite{billardexclu}, depending on the value of the unknown WIMP-nucleon cross section. In \cite{DanielAlb}, a study has been performed on the capability of directional detectors to probe neutralino dark matter in the Minimal Supersymmetric Standard Model and the Next-to-Minimal Supersymmetric Standard Model with parameters defined at the weak scale. It shows that directional detectors such as the future MIMAC detector (50 m$^3$) will probe spin dependent dark matter scattering on nucleons that are beyond the reach of current spin independent detectors since the scalar and axial cross section are not correlated \cite{DanielAlb}.

\begin{figure}[h!]
\begin{center}
\includegraphics[scale=0.17]{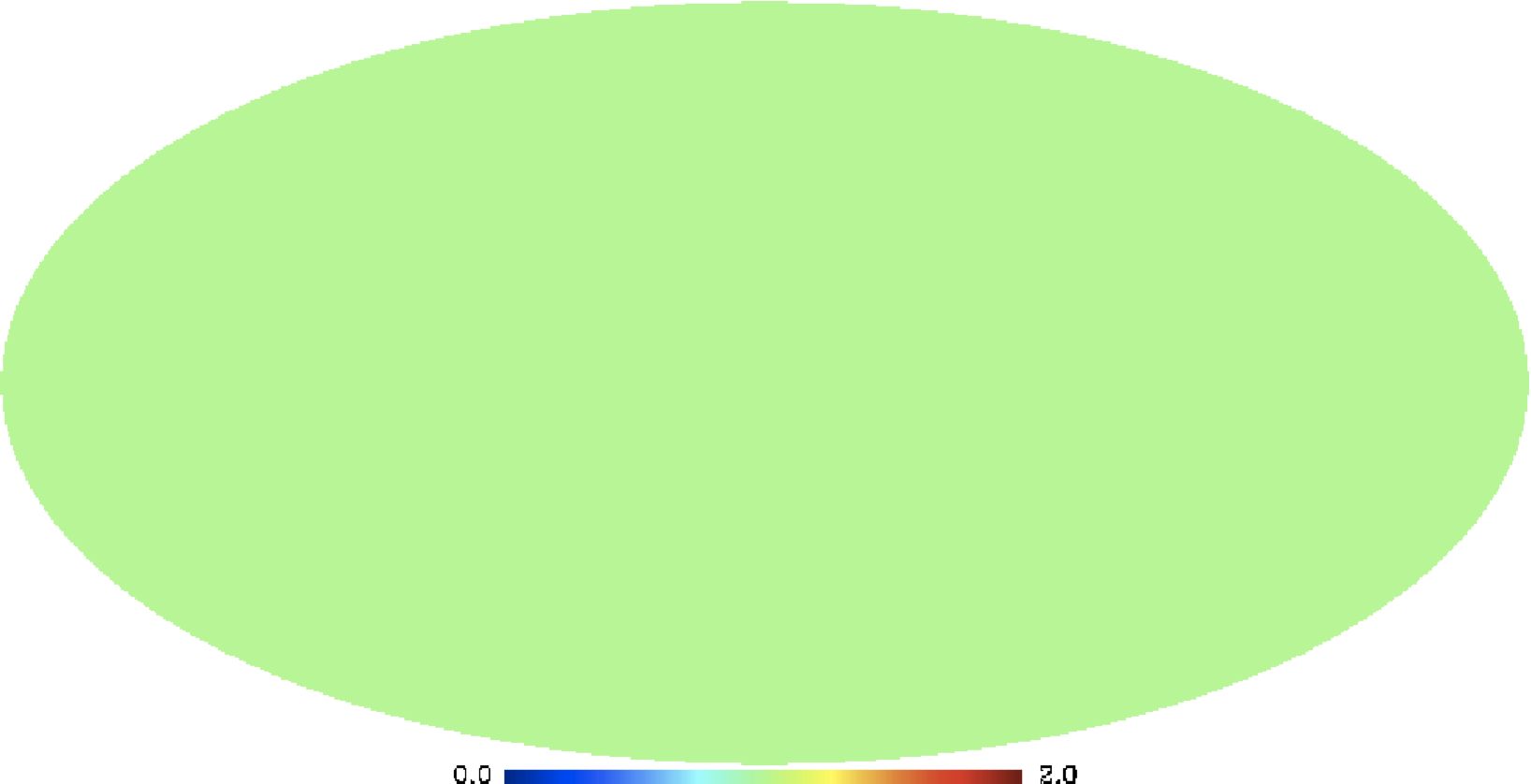}
\includegraphics[scale=0.19,angle=90]{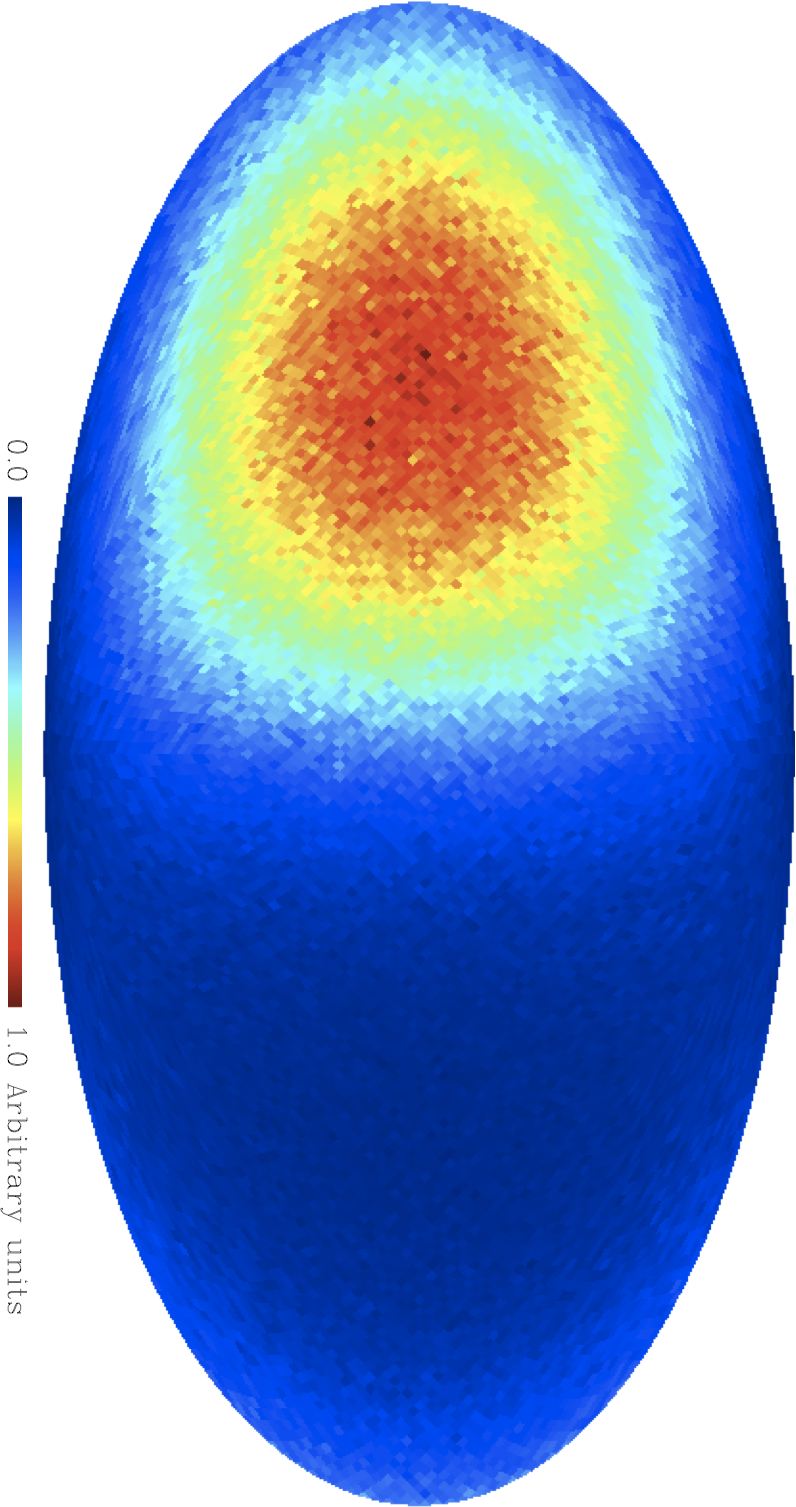}
\includegraphics[scale=0.19,angle=90]{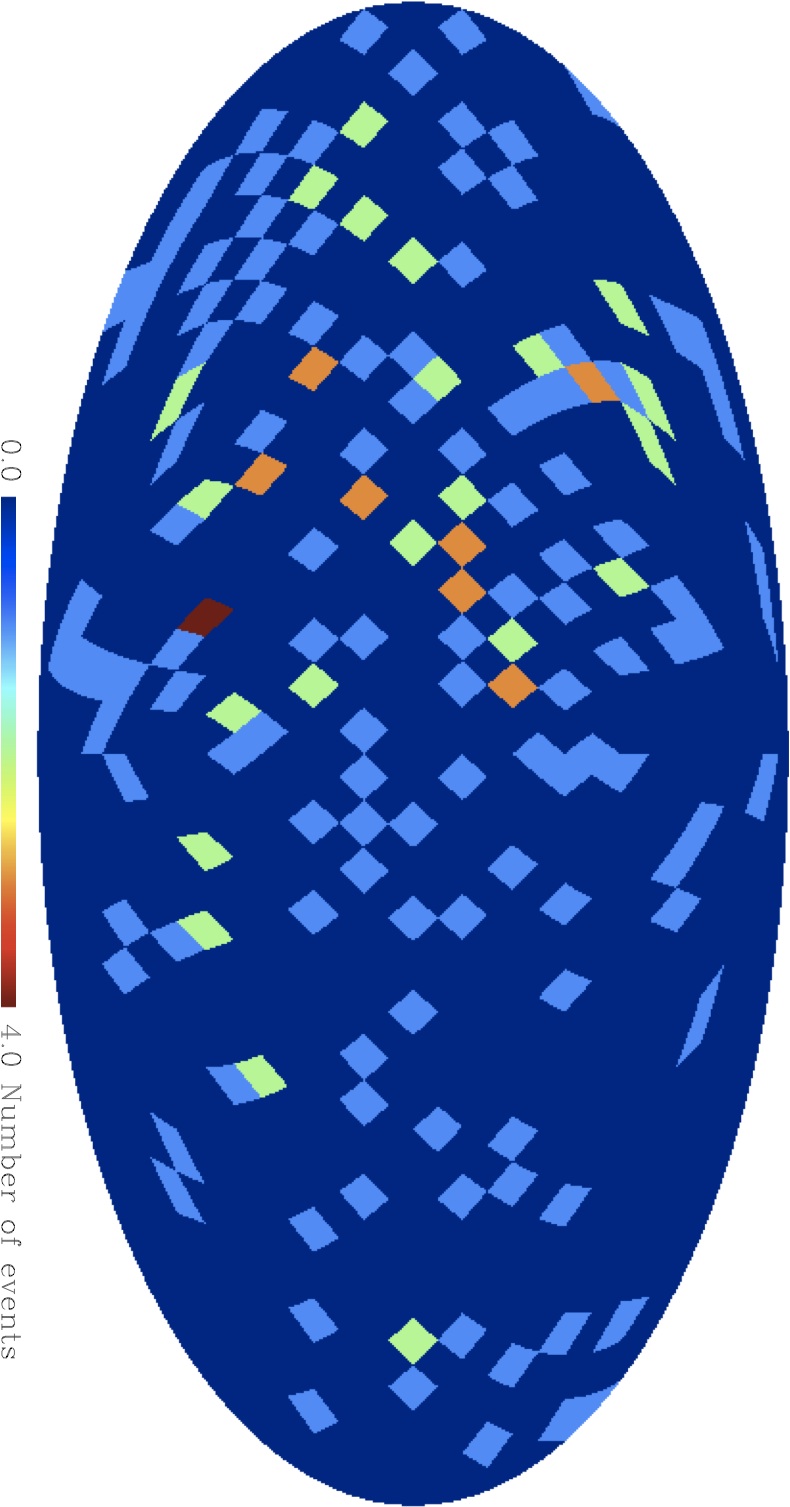}
\caption{From left to right: isotropic background distribution, WIMP-induced 
recoil distribution  in the case of an isothermal spherical halo  and a typical simulated measurement :  
100 WIMP-induced recoils and 100 background events with a low angular resolution. Recoils maps are produced for a $^{19}$F target, a 
 100 GeV.c$^{-2}$ WIMP  and considering recoil energies in the range  5 keV $\leq E_R \leq$ 50 keV. Figures from \cite{billarddisco}.}  
\label{fig:DistribRecul}
\end{center}
\end{figure}

The right panel of figure \ref{fig:DistribRecul}  presents a typical recoil distribution observed by a directional detector : $100$ WIMP-induced events and $100$ background events generated isotropically.  
For an elastic axial cross-section on nucleon $\rm \sigma_{n} = 1.5 \times 10^{-3} \ pb$ and a $\rm 100 \ GeV.c^{-2}$ WIMP mass, this corresponds to an exposure of
% $\rm \sim 7\times 10^3  \ kg.day$ in  $\rm ^{3}He$ and 
$\rm \sim 1.6 \times 10^3 \ kg.day$  in CF$_4$, on their equivalent energy ranges as discussed in ref. \cite{billarddisco}.
  Low resolution maps are used in this case ($N_{\rm pixels} = 768$) which is sufficient  for the low  angular resolution, $\sim 15^\circ$ (FWHM), expected for this type of detector. In this case, 3D readout and sense recognition are considered, while background rejection  is based on electron/recoil discrimination by track length and energy  selection  \cite{Bi-discri}.
In order  to conclude from the recoil map of figure \ref{fig:DistribRecul} (right) that it does contain a fraction of WIMP events pointing towards the direction of the solar motion a likelihood analysis has been developed.
The likelihood value is estimated using a binned map of the overall sky with  Poisson statistics,  as shown in  \cite{billarddisco}.
This is a four parameter likelihood analysis with $m_\chi$,  $\lambda = S/(B+S)$ the  WIMP fraction ($B$ is the  background spatial distribution taken as isotropic and $S$ is the WIMP-induced recoil distribution) and the coordinates ($\ell$, $b$) referring to the maximum of the WIMP event angular distribution.
The result of this  map-based likelihood method is that the main recoil direction is recovered and  is  pointing towards ($ \ell = 95^{\circ} \pm 10^{\circ}, b = -6^{\circ} \pm 10^{\circ}$) at $68 \  \%$ CL, corresponding to a non-ambiguous detection of particles from the galactic halo. This is indeed the discovery  proof of this detection strategy  \cite{billarddisco}.
As emphasized in ref. \cite{Bi-asses}, a directional detector could allow for a high 
significance discovery of galactic Dark Matter even with a small amount of  background contamination.
This holds true even when astrophysical and experimental uncertainties are taking into account.  For very low exposures, competitive exclusion limits may also be imposed \cite{billardexclu}. 

%Directional detection of Dark Matter is based on the fact that the solar system moves with respect to the center of our galaxy with a mean velocity of roughly 220 km/s \cite{spergel}. Taking into account the hypothesis of the existence of a galactic halo of DM formed by WIMPs (Weakly Interacting Particles) with a negligible rotation velocity, we can expect a privileged direction for the nuclear recoils in our detector, coming out from elastic collision with those WIMPs. It has been shown recently, \cite{DarkDisk} that the existence of such dark disk co-rotation is not a problem for this signature.

There are many projects around the world \cite{Drift}, \cite{mit}, \cite{newage}, trying to show the ability to get the directionality at low nuclear recoil energies summarized in \cite{white}.
The MIMAC (MIcro-tpc MAtrix of Chambers) detector project \cite{MIMAC} tries to get these elusive events by a double detection: recoil ionization and track, at low gas pressure with low mass target nuclei (H, $^{19}$F or $^3$He). In order to have a significant cross section we explore the axial, spin dependent, interaction on odd nuclei. The very weak correlation between the neutralino-nucleon scalar cross section and the axial one, as it was shown in 
\cite{PLB}, \cite{DanielAlb} makes this research, at the same time, complementary to the massive target experiments.

\section{ The MIMAC  bi-chamber prototype}

The MIMAC bi-chamber  prototype consists of two chambers of  (10 cm x 10 cm x 25 cm) with a common cathode, which is an elementary module of the future matrix. The purpose of this prototype is to show the ionization and track measurement performances needed to achieve the directional detection strategy.
The primary electron-ion pairs produced by a nuclear recoil in one chamber of the matrix are detected by driving the electrons to the grid of a bulk micromegas \cite{bulk} and producing the avalanche in a very thin gap (256$\mu$m).

\begin{figure}[h!]
\begin{center}
\includegraphics[scale=0.5]{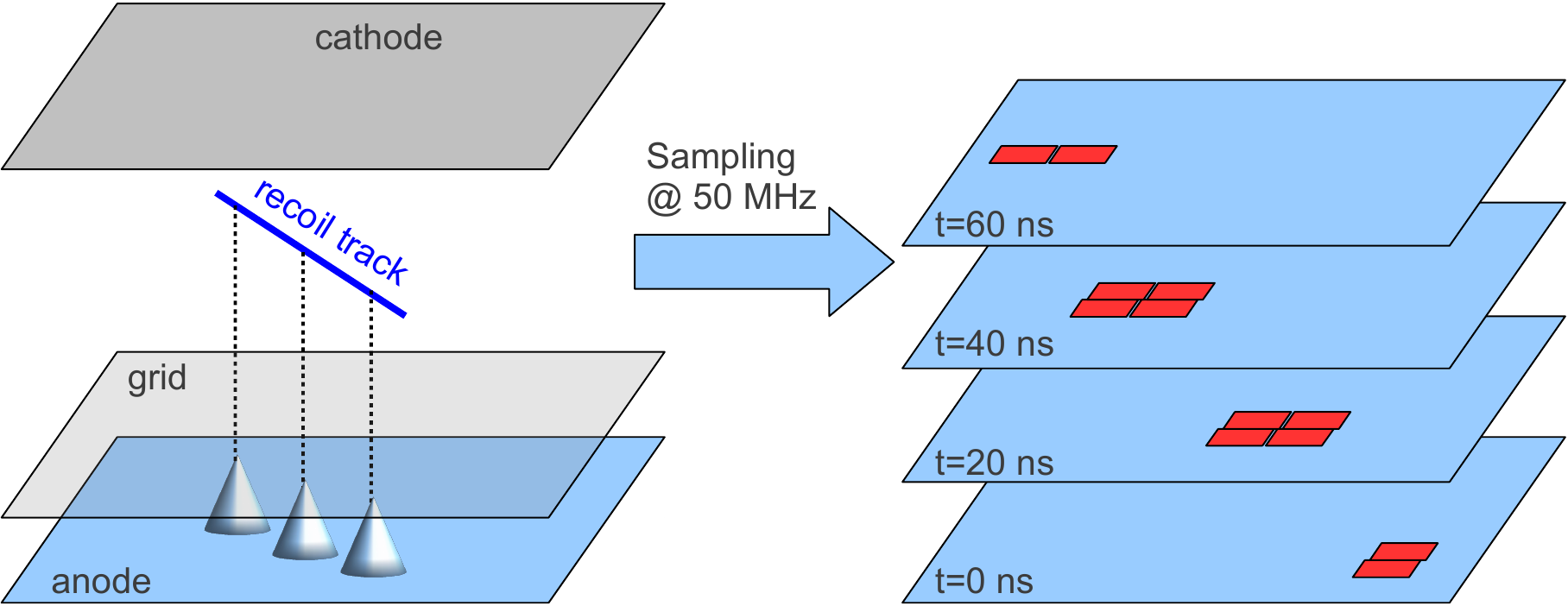}
\caption{The anode is read every 20 ns and 
 knowing the drift velocity of primary electrons,  the 3D track can be reconstructed from the consecutive number of images defining the event.}
\label{recon}
\end{center}
\end{figure}

As pictured on figure  \ref{recon}, the electrons are collected towards the grid in the drift space and are multiplied by avalanche to the pixellized anode thus allowing to get 
information on X and Y coordinates.
To have access to the X and Y coordinates a bulk micromegas  with a 10 by 10 cm$^²$ active area, segmented in pixels with a pitch of 424 $\mu$m was used as 2D readout \cite{Paco}.
 In order to reconstruct the third coordinate Z of the points of the recoil track, the LPSC developed a self-triggered electronics able to perform the anode sampling at a frequency of 50 MHz.
This includes a dedicated 64 channels ASIC \cite{richer} associated to a DAQ \cite{bourrion}.

In order to get the total recoil energy we need to know the ionization quenching factor (IQF) of the nuclear recoil in the gas used. We have developed at the LPSC a dedicated experimental facility to measure such IQF. A precise assessment of the available ionization energy has been performed in $^4$He + 5$\% \rm C_4H_{10}$ mixture within the dark matter energy range (between 1 and 50 keV) by a  measurement of the IQF   \cite{santosQuenching}.
For a given energy, an electron track in a low pressure micro-TPC is an order of magnitude longer and showing more straggling than a recoil one. It opens the possibility to discriminate electrons from nuclei recoils by using both energy and track  information, as it was shown in \cite{DS_cygnus11} and  \cite{Bi-discri}. 

\section{Reconstruction of 3D particle tracks }

The first result concerning the ability to detect tracks with the prototype was performed with a $\rm ^{55}Fe$ X-ray source in order to reconstruct the  5.9 keV electron tracks produced by photoelectric effect in the active volume.
The 3D tracks are obtained from consecutive read-outs of the anode, every 20 ns, defining the event. To get the length and the orientation of the track, a statistical treatment of maximization of the likelihood with the MIMAC observables and an independent measurement of the drift velocity is needed \cite{billardtrack}. Measurements of the drift velocity have been performed \cite{Bi-drift},  and they fit well with Magboltz simulations \cite{Magb}. 

\begin{figure}[h!]
\begin{center}
\includegraphics[scale=0.5]{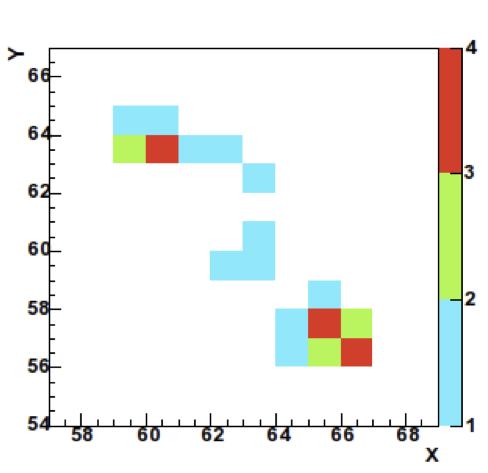}
\includegraphics[scale=0.25]{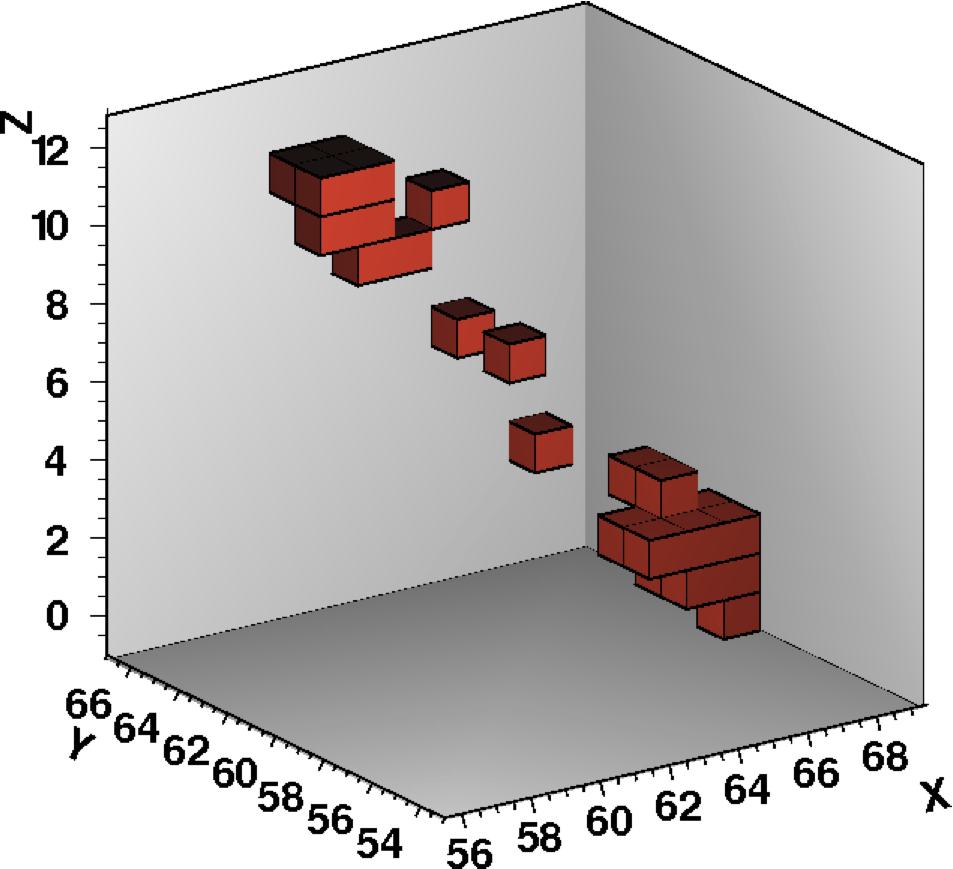}
\includegraphics[scale=0.5]{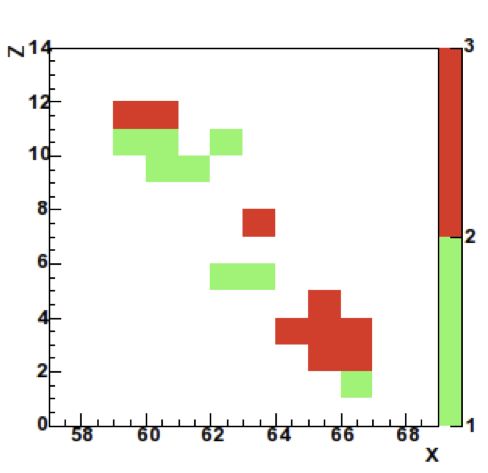}
 \caption{A 5.9 keV electron track in 350 mbar 95\%~$\rm ^4He+C_4H_{10}$.  The left panel represents the 2D projection of the recoil seen by the anode, the center panel represents a 3D view of the track after using the reconstruction algorithm and the right panel represents a projection of the 3D track on the XZ plane}
 \label{electronTrack}
\end{center}
\end{figure}

Figure \ref{electronTrack} presents a typical electron track seen by the anode (X,Y) (left panel), its projection on the XZ plane (right panel) and reconstructed in 3D (center panel).
This result shows the MIMAC capability to reconstruct the track of low energy electrons at high gain which are the  typical background in dark matter experiments.

\begin{figure}[h!]
\begin{center}
\includegraphics[scale=0.25]{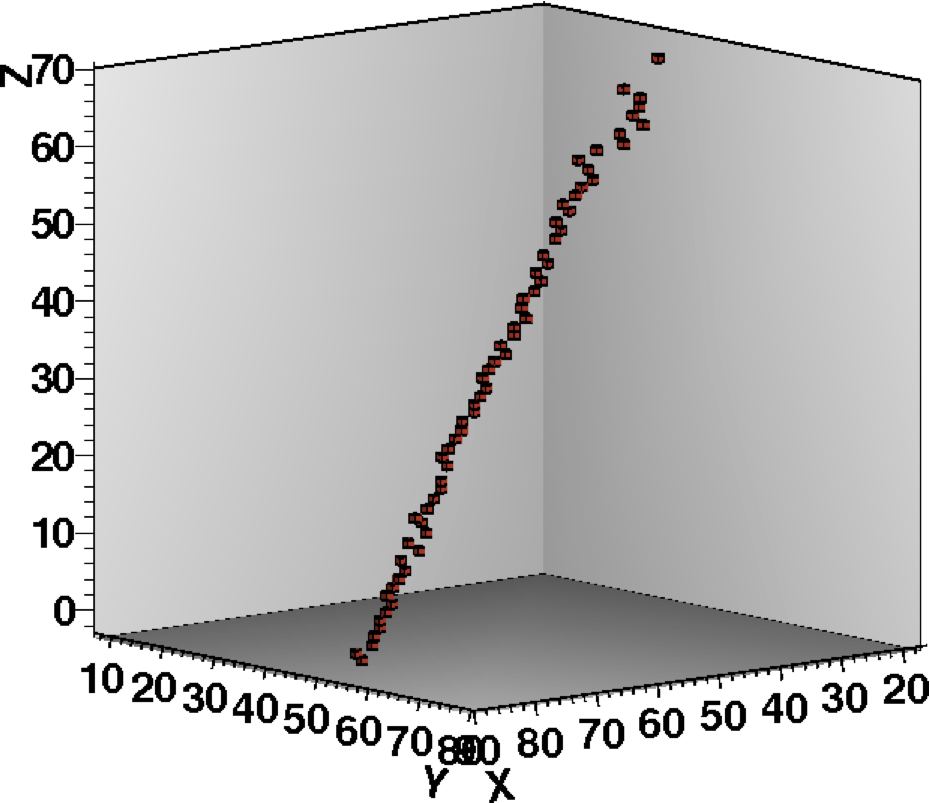}
\includegraphics[scale=0.25]{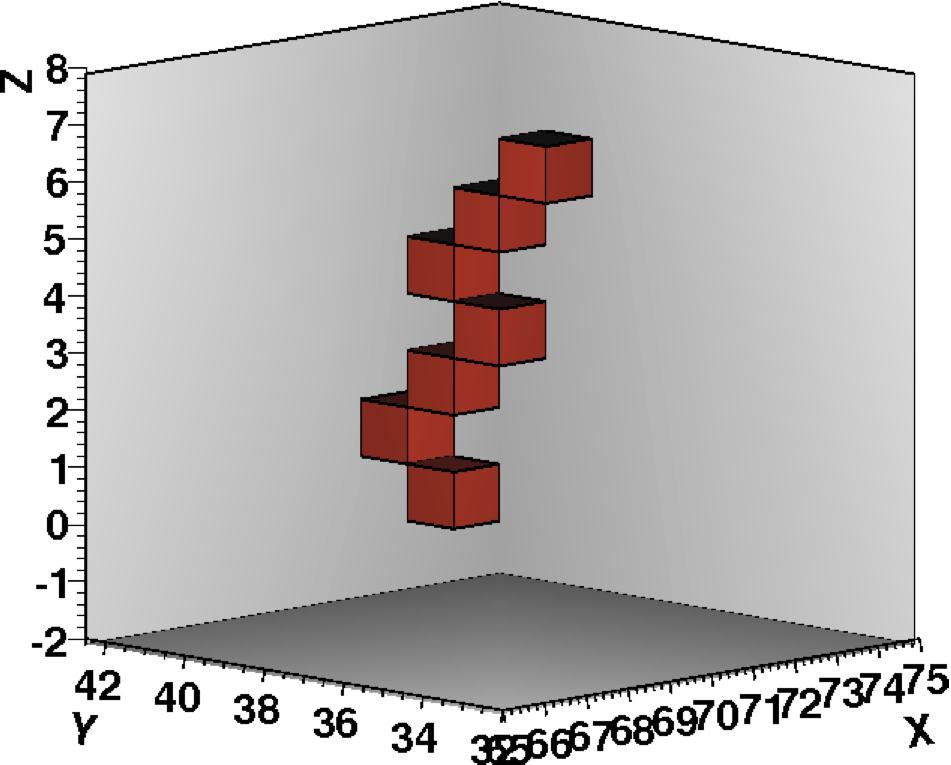}
\includegraphics[scale=0.25]{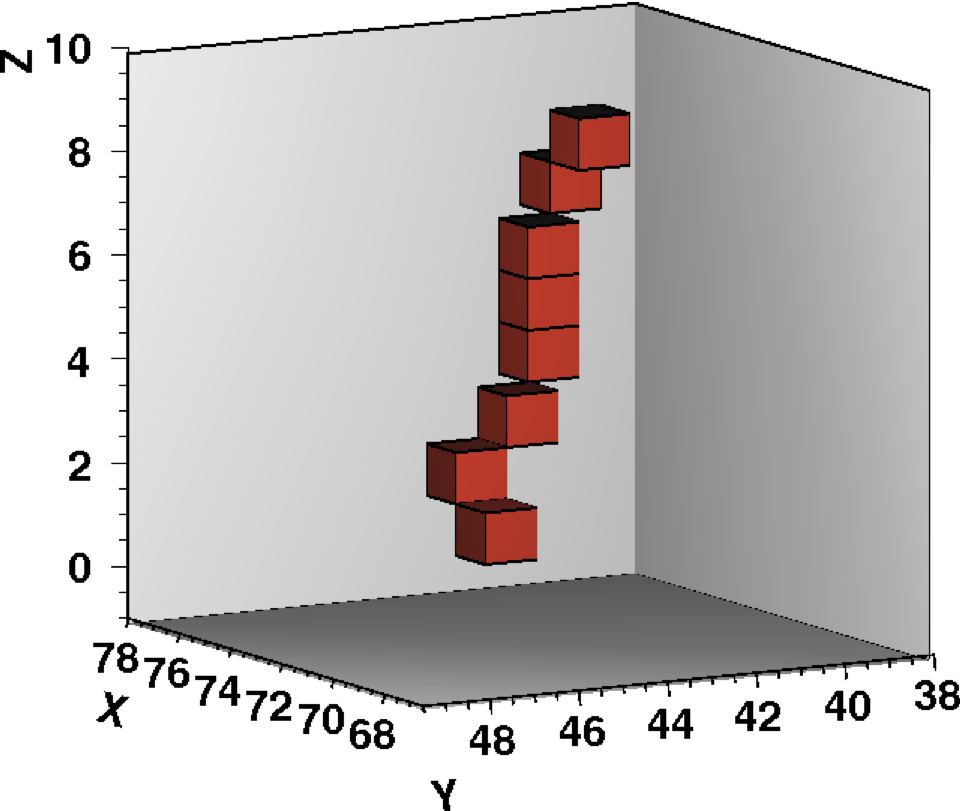}
\caption{From left to right: 3D track reconstructions, from the barycenters of the anode strips of pixels read every 20 ns for : a 5.5 MeV  alpha particle in 350 mbar  of $\rm ^4He+ 5\% C_4H_{10}$, a  8 keV hydrogen nucleus in 350 mbar of $\rm ^4He+ 5\% C_4H_{10}$ and a fluorine nucleus leaving 50 keV in ionization in 55 mbar 70\% $\rm CF_4$ + 30\% $\rm CHF_3$}
\label{nucleiTracks}
\end{center}
\end{figure}

On fig \ref{nucleiTracks} (left panel),  a 3D track reconstruction is achieved for high energy (5.5 MeV) alpha particles issued from the natural radioactivity ($\rm ^{222}Rn$) present in the chamber,  from the barycenters of the anode strips of pixels read every 20 ns.
However, the final validation concerning the possibility for MIMAC to get directional detection had to be done with neutrons giving nuclear recoils in the range of a few keV. In order to have mono-energetic neutron fields, in the range of a few tens of keV, we have performed an experiment at the AMANDE facility (IRSN- Cadarache) allowing to select the energy of the neutrons by the angle with respect to a proton beam producing a neutron resonance on a LiF target.

On fig \ref{nucleiTracks} (center and right panel), 3D tracks of nuclear recoils following elastic scattering of mono-energetic neutrons are represented.
On the center panel, a 8 keV proton recoil leaving a track of 2.4 mm long in 350 mbar of $\rm ^4He+ 5\% C_4H_{10}$ is represented. 
The right panel presents a 50 keV (in ionization) fluorine recoil of 3 mm long obtained in a 55 mbar mixture of 70\% $\rm CF_4$ + 30\% $\rm CHF_3$. 

%In addition, the electron-recoil discrimination, a very important point for dark matter detection, showing the ability to separate  gamma background from nuclear recoils, has been presented in pure Isobutane or $\rm ^4He+ 5\% C_4H_{10}$ mixture in ref. \cite{DS_cygnus11}.

\section{MIMAC at the Underground Laboratory of  Modane (LSM) }

\begin{figure}[h!]
\begin{center}
\includegraphics[scale=0.8]{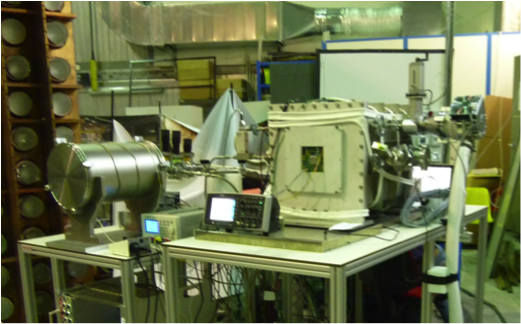}
 \caption{ The bi-chamber module  installed at Modane in June 2012.}
\label{Inst_Modane}
%\end{minipage}
\end{center}
\end{figure}

In June 2012, we have installed the bi-chamber module, at the Underground Laboratory of Modane (LSM), see fig.\ref{Inst_Modane} . 

\begin{figure}[h]
\begin{center}

\includegraphics[scale=0.22]{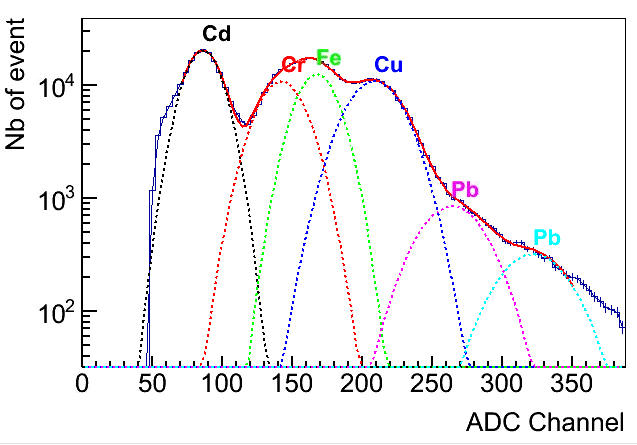}
\includegraphics[scale=0.22]{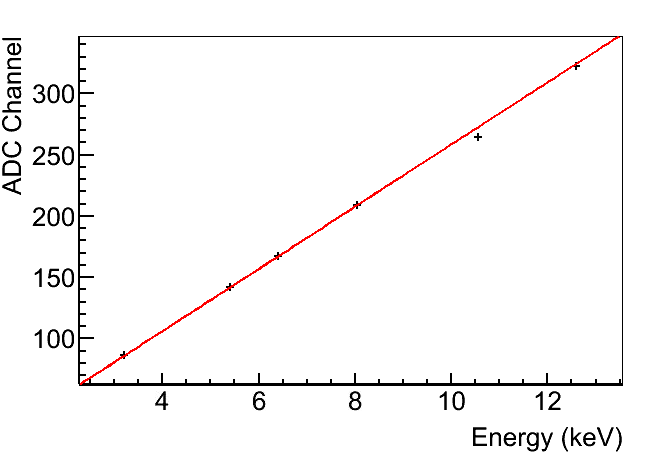}
\includegraphics[scale=0.72]{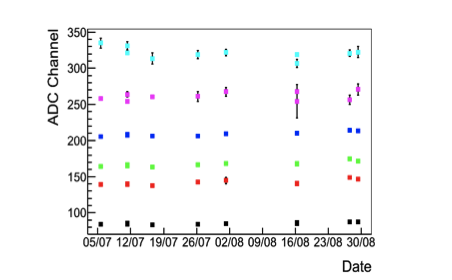}
\caption{Left: The fit of the five X-ray peaks produced by fluorescence. Centre: The linear calibration obtained from the fit. Right:  the gain stability from July 5th to August 28th.}
\label{Modane}
\end{center}
\end{figure}

The set-up includes a close loop gas system with in-line filtering and is able to reach a vacuum of 10$^{-6}$ mbar.  The pressure was regulated at 50 mbar. The gas mixture used was  70\% $\rm CF_4$ + 28\% $\rm CHF_3$ +  2\% $\rm C_4H_{10}$ We have calibrated twice a week by means of fluorescence produced by X generator on thin foils of Cd (3.2 keV), Fe (6.4 keV), Cu (8.1 keV) and Pb (10.5 and 12 keV). In the fig. \ref{Modane},  we show the low energy calibration obtained, its linearity and the stability of the calibration given by the bin position of the different peaks as a function of time. The first data acquisition started on June 22$^{nd}$ and it has been  continuously run and remotely monitored up to October 12$^{th}$ . 

In order to characterize the total background of our detector at Modane, we worked without any shielding. Besides the very good stability of the calibration validating the gas circulation, one of the most interesting facts observed during this first run was the determination of the main sources of $^{222}$Rn inside the gas circuit. 

\begin{figure}[h]
\begin{center}
\includegraphics[scale=0.55]{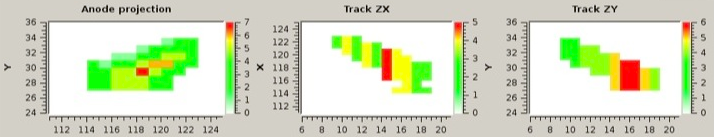}
\caption{On the left, the X-Y plane of the anode showing the intersections of the strips fired. On the centre, the X projection as a function of time, every 20 ns. On the right, the same but for the Y projection}
\label{recoilevent}
\end{center}
\end{figure}

 The 10 cm by 10 cm micromegas coupled to the MIMAC electronics running the 512 channels  per chamber and working at high gain and without any problem during a long period of time (4 months in our first run at Modane), is in fact the validation of the feasibility of a large TPC for directional detection.
As an illustration of the quality of data obtained at Modane, we show on fig.\ref{recoilevent} a very interesting recoil event of 34 keV in ionization. The analysis of this first run at Modane is in progress.

\section{Conclusions}

Directional detection is a promising search strategy to discover galactic dark matter.
The MIMAC detector provides the energy of a recoiling nucleus and the reconstruction of its 3D track. 
The first 3D tracks  observed with the MIMAC prototype were shown: 5.9 keV electrons  and low energy proton and fluorine recoils.The installation of the first bi-chamber prototype at Modane was succesfully done in June 2012. The stability of the gain was characterized and the main source of background events was determined.
The next step will be to build a demonstrator of 1 m$^3$ to show that a large micro-tpc matrix for directional detection of dark matter search is possible.

%\section{Acknowledgements}

%\ack{The MIMAC collaboration acknowledges the ANR-07-BLANC-0255-03 funding.}


\section{References}
\begin{thebibliography}{99}
\bibitem{spergel}D.~N.~Spergel,  Phys.\ Rev.\  D {\bf 37} (1988) 1353.
\bibitem{Bi-asses} J. Billard, F. Mayet and D. Santos, Phys. Rev. D {\bf 85} (2012) 035006
\bibitem{billardtrack} J.~Billard {\it et al.}, JCAP  {\bf 04} (2012) 006.
\bibitem{billarddisco} J.~Billard {\it et al.}, Phys. Lett. B {\bf 691} (2010) 156-162. %arXiv:0911.4086
\bibitem{billardexclu} J.~Billard {\it et al.}, Phys. Rev. D {\bf 82} (2010)  055011.
\bibitem{PLB} E.~Moulin, F.~Mayet and D. ~Santos. Phys.\ Lett. B {\bf 614} (2005) 143-154.
\bibitem{DanielAlb} D. ~Albornoz Vazquez  {\it et al.} Phys. Rev. D {\bf 85} (2012)  055023.
\bibitem{bulk}I. Giomataris {\it et al.}, Nucl. Instrum. Methods A {\bf 560} (2006) 405.
\bibitem{Paco}F. ~Iguaz  {\it et al.}, JINST 6 (2011) P07002 
\bibitem{santosQuenching} D. Santos {\it et al.}, arXiv:0810.1137.
\bibitem{DS_cygnus11}D. Santos {\it et al.} , CYGNUS 2011,  EAS Publications Series 53 (2012) 25-31.
\bibitem{Bi-discri}J. Billard  {\it et al.},  JCAP 07 (2012) 020.
%\bibitem{grignonMPGD} C.~Grignon {\it et al.}, JINST 4 (2009) P11003, arXiv:0909.0654.
\bibitem{Magb} S.F. Biagi, Nucl. Instrum. and Meth. A{\bf 421} (1999) 234-240. 
\bibitem{Bi-drift} J. Billard  {\it et al.}, in preparation.
\bibitem{richer} J.P. Richer {\it et al.}, Nucl. Instrum. Methods A  {\bf 620} (2010) 470.
%arxiv 0912.0186 
\bibitem{bourrion} O. Bourrion {\it et al.},Nucl. Instrum. and Meth. A{\bf 662}(2010) 207.
%\bibitem{xenon}E.~Aprile {\it et al.}, arXiv:1005.0380. 
%\bibitem{cdms}Z. Ahmed {\it et al.}, Science  327 (2010) 1619     
%\bibitem{edelweiss-armengaud}E. Armengaud {\it et al.}, Phys.\ Lett.\  B {\bf 687} (2010) 294-298. 


%\bibitem{agreen}A. M. Green \&  B. Morgan, {\em Astropart.\ Phys.}\  {\bf 27}, 142 (2007).

\bibitem{MIMAC}D.~Santos {\it et al.}, J.\ Phys.\ Conf.\ Ser.\  {\bf 65} (2007) 012012.
\bibitem{Drift}G.~J.~Alner {\it et al.}, Nucl.\ Instr.\ Meth.\  A {\bf 555} (2005) 173.
\bibitem{mit}S.~Ahlen {\it et al.}, arXiv:1006.2928.
\bibitem{newage}K.~Miuchi {\it et al.}, Phys.\ Lett.\  B {\bf 654} (2007) 58.
\bibitem{white}S.~Ahlen {\it et al.}, Int.\ J.\ Mod.\ Phys.\  A {\bf 25} (2010) 1.




\end{thebibliography}
\end{document}